\documentclass[usenatbib]{mn2e}

\input epsf

\usepackage{graphicx}
\usepackage{txfonts}
\usepackage{epsfig}

\title[Estimation of p-mode frequencies]{A devil in the detail:
  parameter cross-talk from the solar cycle and estimation of solar
  p-mode frequencies}

\author[Chaplin et al.]{W.~J.~Chaplin$^1$\thanks{E-mail:
w.j.chaplin@bham.ac.uk}, S.~J.~Jim\'enez-Reyes$^{1,2}$\thanks{E-mail:
sjimenez@iac.es}, A.~Eff-Darwich$^{3,2}$, Y.~Elsworth$^1$,
R.~New$^4$\\ $^1$ School of Physics and Astronomy, University of
Birmingham, Edgbaston, Birmingham, B15 2TT\\ $^2$ Instituto de
astrof\'\i sica de Canarias, 38205, La Laguna, Tenerife, Spain\\ $^3$
Dept. Edafolog\'ia Geolog\'ia, Universidad de La Laguna, E-38205,
Tenerife, Spain\\ $^4$ Faculty of Arts, Computing, Engineering and
Sciences, Sheffield Hallam University, Sheffield S1 1WB\\}

\begin{document}

\maketitle

\begin{abstract}

Frequencies, powers and damping rates of the solar p modes are all
observed to vary over the 11-yr solar activity cycle. Here, we show
that simultaneous variations of these parameters give rise to a subtle
cross-talk effect, which we call the ``devil in the detail'', that
biases p-mode frequencies estimated from analysis of long power
frequency spectra. We also show that the resonant peaks observed in
the power frequency spectra show small distortions due to the effect.
Most of our paper is devoted to a study of the effect for
Sun-as-a-star observations of the low-$l$ p modes. We show that for
these data the significance of the effect is marginal. We also touch
briefly on the likely $l$ dependence of the effect, and discuss the
implications of these results for solar structure inversions.

\end{abstract}


 \section{Introduction}
 \label{sec:intro}

Long timebase observations of the ``Sun as a star'' made, for example,
by the ground-based BiSON network (Chaplin et al. 2007) and the GOLF
(Gabriel et al. 1995) and VIRGO/SPM (Fr\"ohlich et al. 1997)
instruments on the \emph{ESA/NASA} SOHO spacecraft, provide key data
on the low-degree (low-$l$) p modes for probing the deep solar
interior and core. This paper is concerned chiefly with aspects of
analysis of these Sun-as-a-star data.

Accurate and precise p-mode parameter data are a vital prerequisite
for making robust inference on the interior structure of the Sun, and
improved precision, and usually also improved accuracy, accrue from use
of long helioseismic datasets (e.g., see discussion in Chaplin \& Basu
2007). However, long helioseismic datasets must by necessity span a
sizable fraction, or more, of an 11-yr solar activity cycle. Mode
frequencies, damping rates, powers and peak asymmetries
(Jim\'enez-Reyes et al 2007) all vary systematically with the solar
cycle, and these variations can give rise to complications for the
analysis and interpretation, with implications for parameter
accuracies, when long datasets are analyzed in one piece.

There is the concern that changes in frequency over a long dataset may
distort the underlying shapes of the mode peaks to such an extent that
the shapes no longer match the Lorentzian-like functions that are used
to model them. This `distortion' effect turns out not to be a
significant cause for concern, at least where the Sun is concerned
(Chaplin et al., 2008), although it may be an issue for stars that
have shorter, and more pronounced, activity cycles than the Sun. There
is, however, another potential problem for the solar observations, a
`cross-talk' effect that has its origins in the simultaneous
variations of the mode frequencies, powers and linewidths over the
solar cycle. This cross-talk effect can bias the estimated frequencies
of the modes. The effect, which we call the ``devil-in-the-detail'',
is the subject of this paper.

To introduce the effect, let us begin with a fictitious scenario, in
which the p-mode frequencies are the only mode parameters that are
observed to vary over the solar cycle. A mode peak observed in the
power frequency spectrum of a long dataset will then have a centroid
frequency that corresponds to the unweighted average of its
time-varying frequency, over the period of observation. This simple
rule of thumb holds because the frequency shifts of the low-$l$ solar
p modes are small, both in fractional terms and compared to the peak
linewidths. In sum: the unweighted time-averaged frequency may be
directly recovered.

Now consider what happens if the power of the mode is also observed to
vary in time. This will mean that timeseries data at different epochs
will have varying contributions to the final peak profile observed in
the power frequency spectrum. Data samples from times when the mode is
more prominent (i.e., at high power) will carry proportionately larger
weights in determining the final peak profile than samples from times
when the mode is less prominent (i.e., at low power). Because the
frequency also varies in time, one would expect the mode peak to be
pulled, or biased, to a lower or higher frequency than the unweighted
average frequency, with the size and sign of the effect dependent on
the comparative time variations of the frequency and power. This is
our devil in the detail.

Why might the devil in the detail matter? First, it raises uncertainty
over what it is we are actually measuring when we determine mode
frequencies from long datasets. If we have two sets of data that are
not contemporaneous, it creates uncertainty if we wish to correct
their frequencies to a common level of activity. Correction
procedures, which attempt to `remove' the solar-cycle effects from the
fitted frequencies, can be quite sophisticated (e.g., the BiSON
correction procedure; see Basu et al. 2007); but they assume that the
observed mode frequencies correspond to unweighted time averages of
the frequencies. The devil-in-the-detail effect means this is not true
in practice. Furthermore, because the frequency bias that is
introduced depends on the relative sizes of the parameter variations
-- the larger the fractional power variation, for a given frequency
variation, the more pronounced the effect -- we might also expect
there to be some dependence of the bias on the angular degree,
$l$. Both issues above have potential implications for structure
inversions made with the estimated p-mode frequencies.

The layout of the rest of the paper is as follows. In
Section~\ref{sec:prob}, we spell out why cross-talk from simultaneous
variations of the mode parameters introduces the devil-in-the-detail
bias in estimation of frequencies from Sun-as-a-star data. We then
describe how artificial data were used to illustrate, and quantify,
the problem.  We give some important background on the Sun-as-a-star
observations in Section~\ref{sec:sas}, and details on the artificial
Sun-as-a-star data in Section~\ref{sec:data}. We then discuss results
from analysis of long (9.5-yr) artificial datasets in
Section~\ref{sec:full}, where we attempt to quantify the likely size
of the devil-in-the-detail effect for real observations.  Finally, we
pull together the main points of the paper in Section~\ref{sec:conc},
where we discuss the implications for analysis of real data. We also
touch on the issue of the $l$ dependence of the effect, and the
possible impact on results of structure inversions.

 \section{The devil in the detail: parameter cross-talk and the solar cycle}
 \label{sec:prob}

To illustrate the devil-in-the-detail effect, and its impact on the
observed mode profiles in the power frequency spectrum, we begin by
making use of Monte Carlo simulations of single artificial p modes. We
consider simulations of full Sun-as-a-star datasets later in
Section~\ref{sec:art}.

The Laplace transform solution of the equation of a forced, damped
harmonic oscillator was used to generate mode components at a 40-s
cadence in the time domain, in the manner described by Chaplin et
al. (1997).  Components were re-excited independently at each sample
with small `kicks' drawn from a Gaussian distribution. The simulations
gave modes having Lorentzian limit shapes in the power frequency
spectrum. Simulation of the data in the time domain allowed us to
introduce small, time-dependent perturbations of the oscillation
parameters to simulate the effects of the solar-cycle parameter
shifts. We consider in this section sets of simulation results where
parameter shifts were introduced as simple linear functions in
time, $t$. The frequencies in each timeseries of length $T$ were
perturbed linearly according to:
 \begin{equation}
 \nu(t)=\nu(0)+\Delta\nu\,t/T,
 \label{eq:linnu}
 \end{equation}
where $\nu(0)$ was the initial frequency of the mode, and the constant
$\Delta\nu$ fixed the full increase in $\nu(t)$ from the
beginning to the end of the timeseries.

The quantity of interest for measuring the prominence of a mode in the
power frequency spectrum is the maximum power spectral density, or
height $H$, of the resonant peak. There are two independent ways in
which the height may be varied in time. First, the energy given per
unit time to the oscillator may be varied, for example by systematic
variation of the standard deviation describing the excitation
`kicks'. This first scenario leads to changes in height, but the
damping rate, and therefore the peak width $\Gamma$, remains fixed. In
the second scenario the damping rate is varied, while the energy given
per unit time to the oscillator remains fixed. This second approach
leads to variation over time of the width \emph{and} the height of the
resonant peak. The fractional variation in the height is twice that in
the width, and the fractional changes have opposite sign. More details
on the oscillator variations may be found in Appendix~\ref{sec:vars}.

On the Sun, an increase in the surface activity brings with it an
increase in the frequency and damping rate, and a decrease in the
height (and total power) of a typical low-$l$ mode.  Indeed, the
observed sizes and signs of the solar-cycle variations of the
low-$l$ mode parameters appear to be consistent with the second
scenario above (Chaplin et al. 2000), where just the damping is
varied, and the underlying excitation is unchanged over time. Here, we
still tested both scenarios to demonstrate that the size of the
devil-in-the-detail bias introduced in the observed mode frequencies
depends on the scenario that is used.

\begin{figure*}
 \centerline
 {\epsfxsize=8.0cm\epsfbox{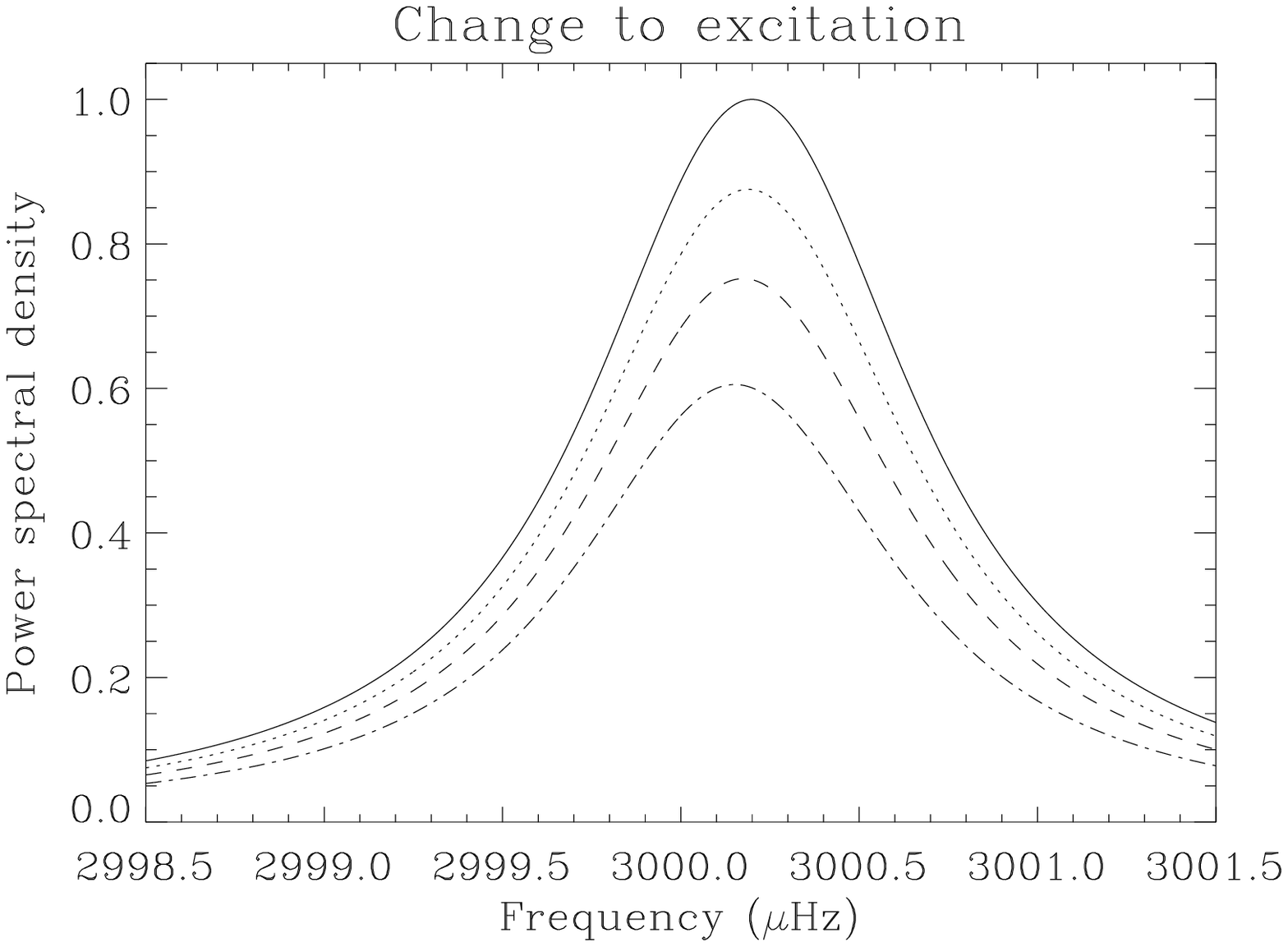}
  \epsfxsize=8.0cm\epsfbox{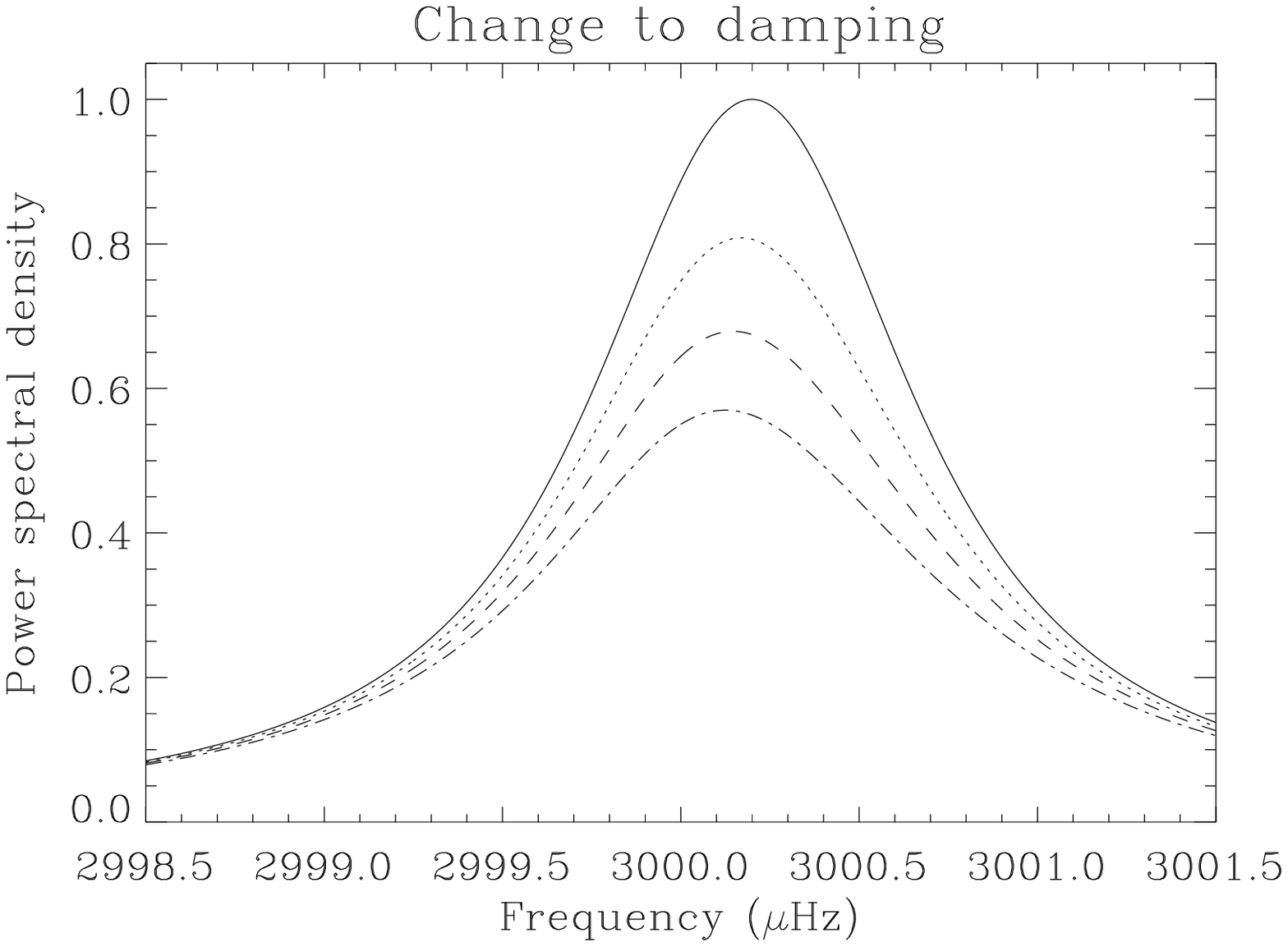}}

 \caption{Results on simulations of single, artificial p modes in
 timeseries where the frequency shift, from start to end, was
 $0.40\,\rm \mu Hz$. Each curve shows the average profile given by
 co-adding the power frequency spectrum of 10,000 independent
 realizations (the co-added spectra were also smoothed with a boxcar
 of width $\sim 0.3\,\rm \mu Hz$). Results in the left-hand panel are
 for the first simulation scenario, where the amplitude of the
 excitation was varied over time; results in the right-hand panel are
 for the second scenario, where the damping rate was varied over time.
 The profile drawn as the solid line in each panel shows the average
 peak for modes which had no height variation over time, i.e., $\Delta
 H = 0$. The other linestyles show results for fractional height
 changes, $\Delta H / H(0)$, of: $-0.25$ (dotted), $-0.50$ (dashed)
 and $-0.80$ (dot-dashed).}

 \label{fig:simlin}
\end{figure*}

In both scenarios variation of the excitation or damping parameter was
contrived in such a way as to give an implied linear variation of the
height over time, i.e.,
 \begin{equation}
 H(t)=H(0)-\Delta H\,t/T.
 \label{eq:linht}
 \end{equation}
In the above, $\Delta H$ describes the full decrease in $H(t)$, from
$t=0$ to $t=T$. We averaged 10,000 independent realizations for
computations made at each chosen combination of $\Delta\nu$ and
$\Delta H$, to recover estimates of the underlying profiles. Some
averaged profiles, which have also been subjected to a small amount of
boxcar smoothing (boxcar width $\sim 0.3\,\rm \mu Hz$), are plotted in
Fig.~\ref{fig:simlin}.  All artificial datasets used to make this plot
simulated a mode having an unperturbed frequency of $\nu(0)=3000\,\rm
\mu Hz$, an unperturbed linewidth of $\Gamma(0) = 1\,\rm \mu Hz$, and
a frequency shift $\Delta\nu = 0.40\,\rm \mu Hz$.

The curves in the left-hand panel show results for simulations made
under the first scenario: here, the amplitude of the excitation was
varied over time. The profile drawn as the solid line in the figure
shows the average peak for modes which had no height variation,
i.e., $\Delta H = 0$. The other linestyles show results for fractional
height changes, $\Delta H/H(0)$, of: $-0.25$ (dotted), $-0.50$
(dashed) and $-0.80$ (dot-dashed) respectively. Note that the dotted
curve has a $\Delta H/H(0)$ very similar to that seen for the most
prominent low-$l$ p modes between solar minimum and maximum. The
curves in the right-hand panel show results for simulations made under
the second scenario: here, the damping rate was varied over time. The
various linestyles show results for the same $\Delta H/H(0)$ as the
left-hand panel.

%%%%%%%%%%%%%%%%%%%%%%%%%%%%%%%%%%%%%%%%%%%%%%%%%%%%%%%%%%%%%%%%%%%%%%%

%%%%%%%%%%%%%%%%%%%%%%%%%%%%%%%%%%%%%%%%%%%%%%%%%%%%%%%%%%%%%%%%%%%%%%%

The curves in Fig.~\ref{fig:simlin} show clear, and predictable,
trends. In the absence of any height variation (solid curves) modes
are shifted in frequency by $\Delta \nu /2$, i.e., the shift is
determined by the (unweighted) average size of the
perturbation. However, when variation in $H$ is introduced (other
linestyles) the mode peaks are shifted in frequency by amounts less
than $\Delta \nu /2$. This is the devil-in-the-detail effect.  In
order to explain the effect, we must remember that as $t$ increased in
our artificial timeseries, the $H$ decreased while at the same time
the frequencies increased. Data samples from times when frequencies
were lower therefore carried a proportionately larger weight in
determining the final peak profiles. We may therefore also say that
the effect arises from cross-talk between the time-varying mode
parameters.

The final profiles are in fact well described by the integral over
time of instantaneous Lorentzian profiles sampled at times $t$, having
the appropriate frequency, height and damping rate for that time,
i.e.,
 \begin{equation}
 \left<P(\nu)\right> = \frac{1}{T}\int\limits_{t=0}^{T}
 \frac{H(t)}{1+\xi^2(t)}\,dt,
 \label{eq:int}
 \end{equation}
where
 \begin{equation}
 \xi(t) = 2 [ \nu - \nu(t)] / \Gamma(t).
 \label{eq:int1}
 \end{equation}
Under the first scenario, where the amplitude of the excitation is
varied but the damping remains fixed (left-hand panel of
Fig.~\ref{fig:simlin}), the frequency of the final profile is well
described by a simple, linear weighted summation of the instantaneous
frequencies, $\nu(t)$, where the weights are proportional to
$H(t)$. That the simple weighting should take this form is clear from
Equation~\ref{eq:int}. However, under the second scenario, where the
damping is varied but the excitation amplitude remains fixed
(right-hand panel of Fig.~\ref{fig:simlin}), this simple relation no
longer holds because the final profile is also affected by the change
in the damping over time (again, see Equation~\ref{eq:int}).

The devil-in-the-detail effect also gives rise to small changes in the
shapes of the peak profiles. First, due to changes in frequency the
widths of the peaks are increased compared to the basic input widths
(first scenario) or the (unweighted) average widths (second scenario).
Second, the peaks are observed to be slightly asymmetric: there is
slightly more power present on the high-frequency side of the peaks
compared to the low-frequency side. The asymmetry increases in size as
$\Delta H/H(0)$ is increased. And for a given $\Delta H/H(0)$, the
asymmetry for the second scenario is larger than the asymmetry for the
first scenario.

How large is this peak asymmetry? Let us define the asymmetry of a
peak to be:
 \begin{equation}
 \left< B \right> = \frac{\left< H \right>_{+\left< \Gamma/2 \right>}
                         -\left< H \right>_{-\left< \Gamma/2 \right>}}
                         {2\left< H \right>}.
 \label{eq:ass}
 \end{equation}
Here $\left< H \right>$ is the height of the peak profile; and $\left<
H \right>_{+\left< \Gamma/2 \right>}$ and $\left< H \right>_{-\left<
\Gamma/2 \right>}$ are the heights of the peak at frequencies plus and
minus half the linewidth. (Equation~\ref{eq:ass} is consistent with
the asymmetry parameter in the well-used Nigam \& Kosovichev (1998)
asymmetric fitting formula, provided the second-order asymmetry terms
in the fitting formula are ignored; see also Toutain, Elsworth \&
Chaplin 2006.) The asymmetry of the final peaks shown in
Fig.~\ref{fig:simlin} rises to 0.2\,per cent (first scenario;
left-hand panel) and 1.4\,per cent (second, right-hand panel),
respectively, when $\Delta H/H(0)=-0.8$. We discuss the expected
contributions to the asymmetries of real p-mode peaks later in
Section~\ref{sec:ass}.

To summarize, significant variation in the height of a mode over time
is expected to give rise to a cross-talk effect with the frequency
parameter. In the absence of such variation, the location in frequency
of the centroid of the mode peak (in the power frequency spectrum)
would in principle give an accurate measure of the unweighted average
mode frequency over the period of observation. However, when
significant variation of the height is present, the peak of the mode
will be pulled to a slightly lower frequency, and it cannot be assumed
that the estimated frequency is the unweighted average frequency. Some
distortion of the underlying peak shape is also expected.

Next, we go on to ask: How large, and how significant, are the effects
for the low-$l$ solar p modes?

 \section{Artificial Sun-as-a-star data}
 \label{sec:art}

In order to quantify the likely impact of the devil-in-the-detail
effect described above on real observations, we generated, and then
analyzed, 40 artificial datasets each of which was designed to mimic
just under 9.5\,yr of full BiSON or GOLF-like Sun-as-a-star Doppler
velocity observations. In order to understand some of the results one
first has to understand the basic characteristics of the Sun-as-a-star
data. We discuss these characteristics next (Section~\ref{sec:sas})
before going on to give more details on the full artificial datasets
in Section~\ref{sec:data}. Those who are very familiar with the
Sun-as-a-star data should be safe skipping Section~\ref{sec:sas}.

 \subsection{Sun-as-a-star observations}
 \label{sec:sas}

Sun-as-a-star observations take an average, over the visible disc, of
either the Doppler velocity or intensity perturbations associated with
the modes. The technique is inherently very stable and shows excellent
sensitivity to the low-$l$, core-penetrating modes.  When the
observations are made from a location in, or close to, the ecliptic
plane the rotation axis of the Sun is always nearly perpendicular to
the line-of-sight direction, and only some of the $2l+1$ components of
the non-radial modes are clearly visible: those having even $l+m$,
where $m$ is the azimuthal order. This is the perspective from which
the Sun-as-a-star BiSON (ground-based network), and GOLF and VIRGO/SPM
(on the SOHO spacecraft at L1), view the Sun.

A set of Sun-as-a-star observations gives a single time series whose
power frequency spectrum contains many closely spaced mode peaks.
Parameter estimation must contend with the fact that within the
non-radial mode multiplets the various $m$ lie in very close frequency
proximity to one another. Suitable models, which seek to describe the
characteristics of the $m$ present, must therefore be fitted to the
components simultaneously. Furthermore, overlap between modes adjacent
in frequency is a cause for concern over much of the low-$l$ spectrum
(in particular at high frequencies, where the mode peaks are wider due
to heavier damping).

Let us consider the impact of the solar activity cycle. The mode
parameters are observed to vary over time, in response to variations
of the near-surface activity. Estimates of the parameters depend not
only on when the observations were made, but also on the frequencies,
and the spatial properties [i.e., the ($l,|m|$)], of the modes.  The
frequencies matter because the mode inertias are frequency dependent,
and the inertias affect the sensitivity of the modes to the
near-surface perturbations. The spatial properties matter because the
distribution of the near-surface activity is spatially
non-homogeneous. The response of the modes to the changing activity is
therefore in part determined by the combination ($l,|m|$). This
spatial dependence, in combination with the visibility characteristics
of the Sun-as-a-star data, creates a complication that has already
been well documented in the literature (e.g., Chaplin et al. 2004). We
raise this complication here because it plays a part in understanding
results on the devil-in-the-detail effect, presented later in
Section~\ref{sec:basic}.

The complication arises for modes with $l \ge 1$, because then some of
the mode components are no longer visible in the Sun-as-a-star data
(see above). Since these components are in effect missing from the
observations, and those components that are present have solar-cycle
frequency shifts that are ($l$,$|m|$)-dependent, it is not possible to
estimate accurately the frequency centroids of the non-radial
modes. It is the centroids that carry information on the spherically
symmetric component of the internal structure, and which are therefore
required as input for the hydrostatic structure inversions. In the
complete absence of the near-surface activity, the missing components
would be an irrelevance. All mode components would then be arranged
symmetrically in frequency, meaning centroids could be estimated
accurately from the subset of components visible to the Sun-as-a-star
data. A near-symmetric arrangement is found at the epochs of the
modern cycle minima. However, when the observations span a period
showing medium to high levels of activity -- as a long dataset by
necessity must -- the frequencies given by fitting the Sun-as-a-star
data differ from the true centroids by an amount that is sensitive to
$l$. This $l$ sensitivity arises because modes of different $l$ are
comprised of visible components having different combinations of $l$
and $|m|$; and these different combinations show different solar-cycle
frequency shifts (in amplitude and phase).

The `complication' outlined above is important for what follows, in
that we first had to quantify the $l$ dependence of the frequencies
introduced by it, in order to properly isolate the devil-in-the-detail
effect. This was done using artificial datasets simulating full
Sun-as-a-star observations. Next, we go on to describe how the
datasets were made.

 \subsection{The full artificial datasets}
 \label{sec:data}

Each of the full artificial datasets had a simulated length of
3456\,days ($\sim 9.5\,\rm yr$), and was constructed component by
component in the time domain (in the manner described for single p
modes in Section~\ref{sec:prob} above). Sets were made with a full
cohort of simulated low-$l$ modes, covering the ranges $0 \le l \le 5$
and $1000 \le \nu \le 5500\,\rm \mu Hz$. The artificial datasets were
given the same underlying parameters as the datasets used for the
first hare-and-hounds exercise of the solarFLAG\footnote{solarFLAG
URL: http://bison.ph.bham.ac.uk/\,\~\,wjc/Research/FLAG.html}
collaboration. All datasets had different realization noise. More
details on how the datasets were made may be found in Chaplin et
al. (2006).

Solar-cycle-like variations were introduced through systematic
variation in simulated time of the oscillator parameters. A high-order
polynomial fit to the disc-averaged solar 10.7-cm radio flux, observed
over the 9.5-yr period beginning 1992 July 24, was used to describe
the globally averaged activity of our artificial Sun. We used known
sensitivities of the p-mode parameters to variations in the 10.7-cm
flux to calibrate the simulated time variations of the oscillator
parameters.

We introduced systematic changes in mode frequency, which had an
appropriate underlying ($l,|m|$) and frequency dependence. This was
accomplished through appropriately scaled variation (see, for example,
Jim\'enez-Reyes et al 2004) of the frequency parameter of each
component. Solar-cycle-like changes were also introduced in the
component damping rates and powers. These changes were given a weak
frequency dependence (as seen in real observations), but no dependence
on ($l,|m|$)\footnote{Only recently (Salabert et al. 2007) has
evidence for an ($l,|m|$)-dependence of the excitation and damping
parameters been uncovered in the low-$l$ modes, and only then at
marginally significant levels.}.

We constructed two sequences of artificial data. Each sequence was
comprised of 20 independent artificial realizations of a 9.5-yr
dataset.  In the first sequence, which we call sequence FLAG\#0,
solar-cycle-like variations were introduced \emph{only} in the mode
frequencies. Datasets in the second sequence, which we call sequence
FLAG\#1, had variations in mode frequency \emph{and} mode damping and
mode power.

\begin{figure}
 \centerline
 {\epsfxsize=8.0cm\epsfbox{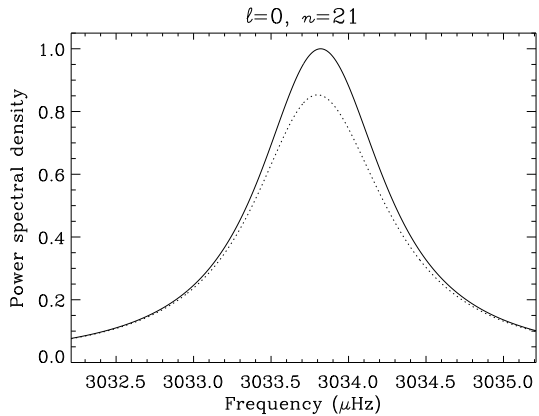}}

 \caption{Final peaks for artificial $l=0$, $n=21$ mode, in
 artificial power spectra made from 9.5-yr of simulated Sun-as-a-star
 oscillations data. The solid line shows the final peak for the
 FLAG\#0 simulation scenario; and the dotted line the final peak for
 the FLAG\#1 scenario (see text).}

 \label{fig:sim3456}
\end{figure}

To illustrate visually the effect given to one of the mode peaks under
these more realistic simulation scenarios, Fig.~\ref{fig:sim3456}
shows profiles for the $l=0$, $n=21$ mode ($\nu(0)=3033.7\,\rm \mu
Hz$). The solid line shows the final peak for the FLAG\#0 scenario;
and the dotted line the final peak for the FLAG\#1 scenario. The
FLAG\#1 peak has an asymmetry of 0.4\,per cent (see also
Fig.~\ref{fig:ass} and discussion at end of Section~\ref{sec:ass}).

%%%%%%%%%%%%%%%%%%%%%%%%%%%%%%%%%%%%%%%%%%%%%%%%%%%%%%%%%%%%%%%%%%%%%%%

%%%%%%%%%%%%%%%%%%%%%%%%%%%%%%%%%%%%%%%%%%%%%%%%%%%%%%%%%%%%%%%%%%%%%%%

\section{Analysis of full 9.5-yr datasets}
\label{sec:full}

\subsection{Overview and basic results}
\label{sec:basic}

The 9.5-yr datasets were analysed in their entirety. This meant we
first computed the power frequency spectrum of each full
dataset. Multi-component models were then fitted to modes in each
power frequency spectrum to extract estimates of the mode parameters.
This was accomplished by maximizing a likelihood function commensurate
with the $\chi^2$, 2-d.o.f. statistics of the power spectral density.
We adopted the usual approach to fitting the Sun-as-a-star spectra
(e.g., see Jim\'enez-Reyes et al 2007), and the low-$l$ modes were
fitted in pairs ($l=0$ with 2, and $l=1$ with 3) in the range $1400
\le \nu \le 3500\,\rm \mu Hz$. This is the range over which it is
possible to extract well-constrained estimates of the solar-cycle
frequency shifts in the real observations. To reduce uncertainties,
the fitted frequencies of each mode were averaged over the 20 datasets
in each sequence, to give two final sets of mean fitted mode
frequencies.

%%%%%%%%%%%%%%%%%%%%%%%%%%%%%%%%%%%%%%%%%%%%%%%%%%%%%%%%%%%%%%%%%%%%%%%

\begin{figure*}
  \centerline{
  \epsfxsize=9.0cm\epsfbox{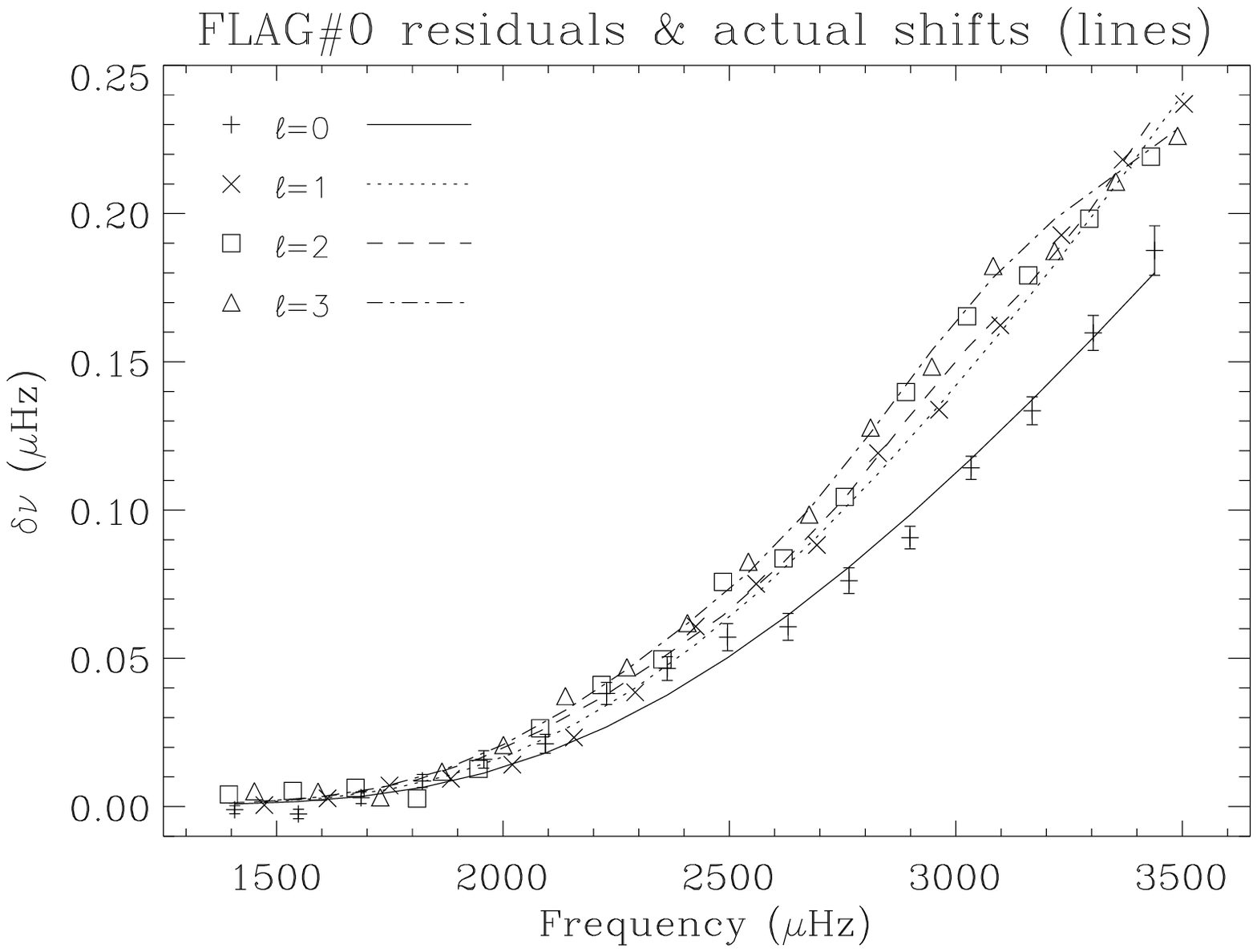} \quad
  \epsfxsize=9.0cm\epsfbox{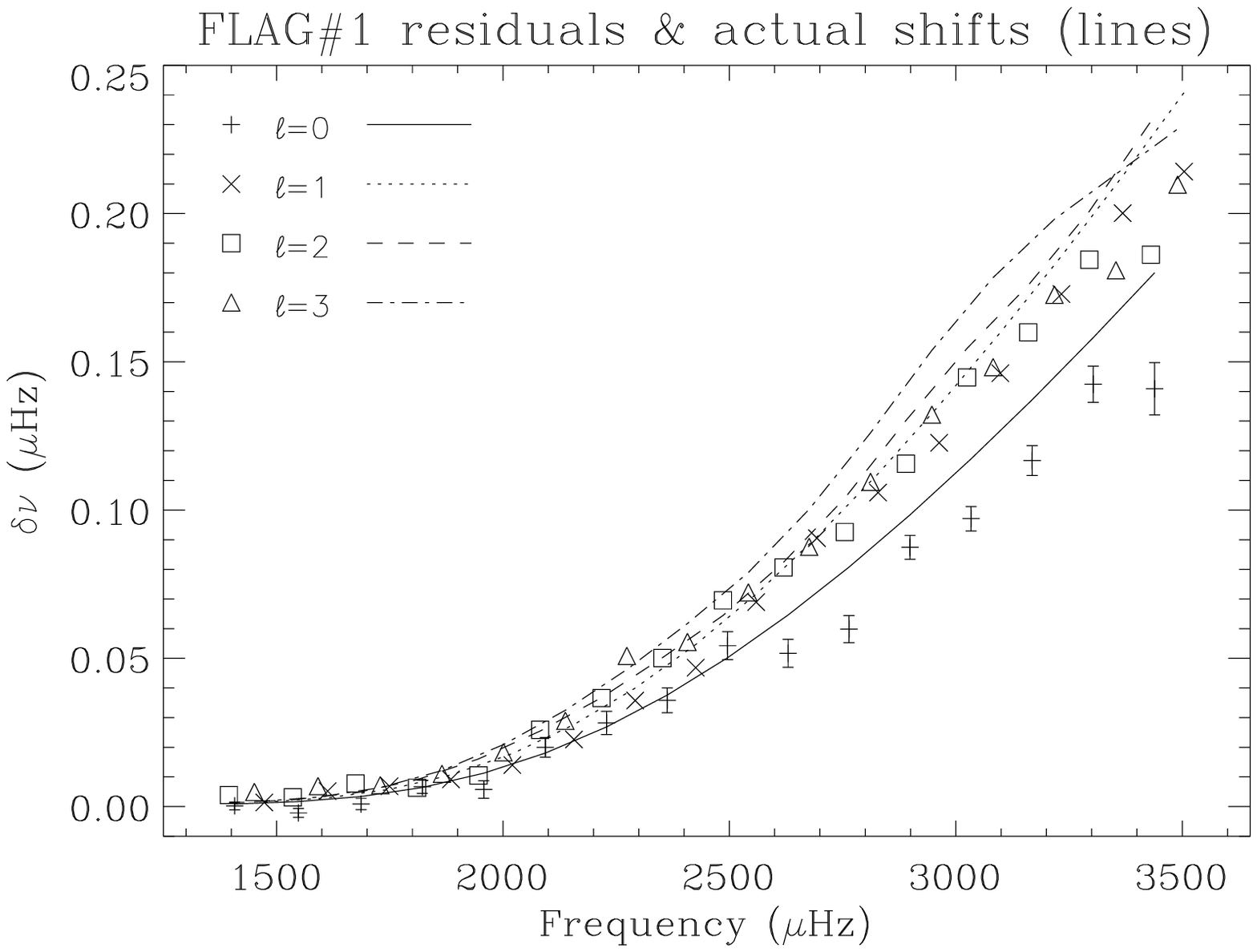}}

  \caption{Main results of the full-spectrum exercises
  (Section~\ref{sec:full}) for the FLAG\#0 (left-hand panel) and
  FLAG\#1 (right-hand panel). Frequency residuals plotted as symbols
  for different $l$ (see annotation) are differences between the
  full-spectrum mean fitted mode frequencies and the unperturbed
  (``zero activity'') input mode frequencies. The residuals are due to
  the solar-cycle frequency shifts introduced into the data. The
  various lines (see annotation) show the unweighted average frequency
  shifts that were introduced in the data (i.e., for a given mode, the
  difference between the unweighted average of the time-varying
  frequency over the 9.5-yr simulated observation, and the unperturbed
  input frequency).}

 \label{fig:full}
\end{figure*}

%%%%%%%%%%%%%%%%%%%%%%%%%%%%%%%%%%%%%%%%%%%%%%%%%%%%%%%%%%%%%%%%%%%%%%%

Fig.~\ref{fig:full} summarizes the main results of this full-spectrum
exercise.  The frequency residuals plotted as symbols for different
$l$ (see annotation) are differences between the mean fitted mode
frequencies and the unperturbed input mode frequencies (with FLAG\#0
differences shown in left-hand panel, and FLAG\#1 differences shown in
the right-hand panel). By unperturbed frequencies, we mean the
underlying input frequencies onto which the solar-cycle-like
variations were added in the artificial data. The unperturbed
frequencies may therefore be regarded as notional ``zero-activity''
frequencies. We have also plotted the errors on the $l=0$ mean fitted
frequencies, to give a guide to the precision in the data
(uncertainties on the mean frequencies at the other $l$ have similar
sizes). The first point to take from the figure is that the mean
fitted frequencies exceed the unperturbed, zero-activity frequencies
by amounts that depend on both frequency and degree, $l$. These
differences are of course due to the solar-cycle frequency shifts that
we introduced in the data.

The various lines in the both panels show the unweighted average
frequency shifts that were actually introduced (see annotation). For
each mode, this shift is just the difference between the unweighted
average of the time-varying frequency over the 9.5-yr simulated
observation, and the unperturbed input frequency. How were these
average frequency shifts calculated? Calculation of the average shifts
is trivial for the $l=0$ and 1 modes, because only one combination of
($l,|m|$) components is observed in the Sun-as-a-star (and hence our
artificial) data at these degrees. The shifts are given by the
unweighted average frequency shifts of the (0,0) and (1,1) components,
respectively, over period of simulated observation, and these are
plotted as solid and dotted lines in both panels of
Fig.~\ref{fig:full}.

At $l=2$ and 3 matters are non-trivial, because of the `complication'
we discussed at the end of Section~\ref{sec:sas} above.  Here, modes
seen in the Sun-as-a-star observations are comprised of components
showing two different combinations of $l$ and $|m|$, each of which has
a different-sized shift over the solar cycle. How should the expected
shifts be computed for these modes?  Chaplin et al. (2004) showed
that the fitted $l=2$ and 3 Sun-as-a-star frequencies may be modelled
as linear combinations of the frequencies of the visible
($l,|m|$). These combinations turn out to be dominated by the
frequencies of the outer $|m|=l$ components, because they are more
prominent than the other visible components in the multiplets. More
recently, Appourchaux \& Chaplin (2007) have given frequency-dependent
estimates of the coefficients that are needed to scale the
combinations at each $l$. In effect, use of the coefficients allows
one to calculate a notional Sun-as-a-star frequency for every mode at
$l=2$ and $l=3$, from an appropriate combination of the frequencies of
the constituent $m$. This we have done here, to give unweighted
Sun-as-a-star averages of the time-varying $l=2$ and 3 frequencies,
from which the unweighted average frequency shifts plotted as dashed
and dot-dashed lines in both panels of Fig.~\ref{fig:full} were
calculated. By applying this procedure, we take care of the $l$
dependence of the frequencies introduced by the missing components,
and therefore our `complication'.

What are the main conclusions to be drawn from the results plotted in
Fig.~\ref{fig:full}? The data plotted in the left-hand panel show that
the fits to the FLAG\#0 data recover the unweighted average frequency
shifts, and therefore the unweighted average frequencies, at the level
of precision of the simulated observations.  Recall that only the
frequencies were varied over time in these datasets. There are, in
contrast, significant discrepancies present in the FLAG\#1
results. This is of course due to the devil-in-the-detail effect. The
FLAG\#1 datasets included within them variations over time of the
heights and widths of the modes. These variations in effect gave a
different weighting to different parts of the timeseries, resulting in
a systematic underestimation of the unweighted average frequencies,
just as was seen in the simpler, linear-shift simulations in
Section~\ref{sec:prob}.

To get a quantitative measure of the cumulative significance of the
differences we did the following. We performed a linear regression of
the FLAG\#0 and FLAG\#1 mean fitted frequency residuals (various
symbols in Fig.~\ref{fig:full}) on the unweighted average frequency
shifts (various linestyles). The data in Table~\ref{tab:full}
correspond to the best-fitting gradients for each $l$. A best-fitting
gradient of unity is expected when the fitted frequencies match the
average input frequencies. Gradients consistent with unity are
returned from analysis of the FLAG\#0 data\footnote{Recall that at
$l=2$ and 3 the unweighted average frequency shifts were computed from
weighted linear combinations of the unweighted average shifts of the
different $|m|$ seen in these modes. The different $|m|$ seen are:
$|m|=2$ and 0 at $l=2$; and $|m|=3$ and 1 at $l=3$.  If the weaker,
inner components are neglected, and only the average shifts of the
outer components are used, the FLAG\#0 best-fitting gradients are
reduced at $l=2$ and 3 to 0.92 and 0.90, respectively. This emphasizes
that allowance must be made for the influence of the weaker, inner
components on the fits when the expected average shifts are
estimated.}. However, for FLAG\#1 all gradients lie significantly
below unity, by anywhere from 5 and $8\sigma$, because the average
input frequencies are systematically underestimated.

%%%%%%%%%%%%%%%%%%%%%%%%%%%%%%%%%%%%%%%%%%%%%%%%%%%%%%%%%%%%%%%%%%%%%%%

\begin{table}

  \caption{Results of performing a linear regression of the FLAG\#0
  and FLAG\#1 fitted frequency residuals on the unweighted average
  frequency shifts present in the datasets. The results in the table
  correspond to the best-fitting gradients.}

  \begin{center}
    \begin{tabular}{lcc}
    \hline
    & \multicolumn{2}{c}{\emph{Best-fitting gradient}}\\
      Degree, $l$ & FLAG\#0& FLAG\#1\\
      \hline
      0....& 1.00$\pm$ 0.02& 0.86$\pm$ 0.02\\
      1....& 1.02$\pm$ 0.01& 0.92$\pm$ 0.02\\
      2....& 1.01$\pm$ 0.02& 0.89$\pm$ 0.02\\
      3....& 0.98$\pm$ 0.02& 0.84$\pm$ 0.02\\
      \hline
    \end{tabular}
  \end{center}
  \label{tab:full}
\end{table}

%%%%%%%%%%%%%%%%%%%%%%%%%%%%%%%%%%%%%%%%%%%%%%%%%%%%%%%%%%%%%%%%%%%%%%%

 \subsection{Size of the devil-in-the-detail effect}
 \label{sec:size}

%%%%%%%%%%%%%%%%%%%%%%%%%%%%%%%%%%%%%%%%%%%%%%%%%%%%%%%%%%%%%%%%%%%%%%%

\begin{figure*}
  \centerline
      {\epsfxsize=14.0cm\epsfbox{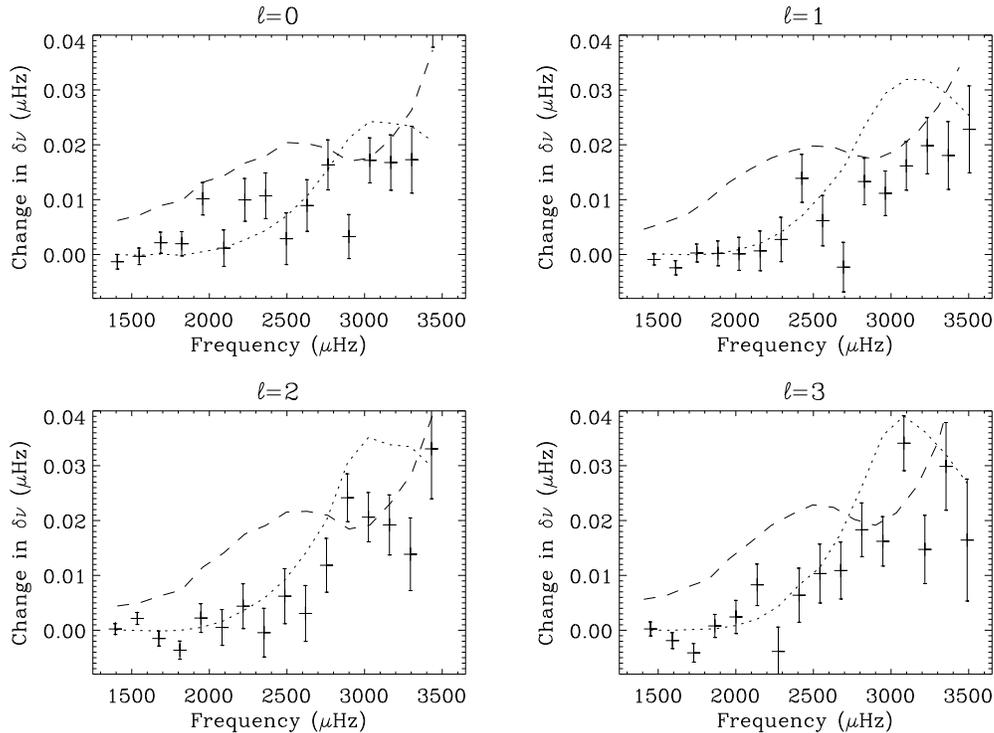}}

  \caption{Dotted lines: sizes of the devil-in-the-detail effect,
 i.e., amounts by which the fitted frequencies are expected to be
 reduced compared to the unweighted average frequencies. Points with
 associated error bars: differences between the mean fitted FLAG\#0
 and FLAG\#1 frequencies. Dashed lines: sizes of the typical frequency
 uncertainties expected from a single 9.5-yr dataset.}

 \label{fig:predcorr}
\end{figure*}

%%%%%%%%%%%%%%%%%%%%%%%%%%%%%%%%%%%%%%%%%%%%%%%%%%%%%%%%%%%%%%%%%%%%%%%

The frequency bias introduced by the devil-in-the-detail effect
corresponds to the amounts by which the fitted frequencies
underestimate the unweighted average frequencies. This bias may
therefore be estimated directly by subtracting the mean fitted FLAG\#1
frequencies from the mean fitted FLAG\#0 frequencies. When we perform
this subtraction, we obtain the bias estimates plotted as crosses with
error bars in Fig.~\ref{fig:predcorr}.

How significant is the bias? We have also plotted in each panel of
this figure the sizes of the typical frequency uncertainties expected
from a single 9.5-yr dataset (dashed lines). The plots show that the
bias for individual modes is typically smaller in size than the
frequency uncertainties. That said, the bias is systematic, and the
cumulative effect is undoubtedly significant (e.g., see data in
Table~\ref{tab:full}).

Finally, we are also in a position to predict the
devil-in-the-detail frequency bias. This may be accomplished
numerically, by using Equation~\ref{eq:int} to construct the expected
final profiles of each mode peak for the known FLAG\#1 input
parameters, and input parameter variations. Comparison of the
frequency maxima of these calculated profiles with the unweighted
average frequencies then gives the expected bias. The resulting
predictions are plotted as dotted lines in each panel of
Fig.~\ref{fig:predcorr}. Fair agreement is seen with the results given
by fitting the artificial data (crosses with errors), although offsets
are present, most notably in the higher frequency range. These offsets
presumably reflect the impact of other subtle bias effects that the
Sun-as-a-star fits are subject to.

 \subsection{Effect on peak shape}
 \label{sec:ass}

\begin{figure}
 \centerline
 {\epsfxsize=8.0cm\epsfbox{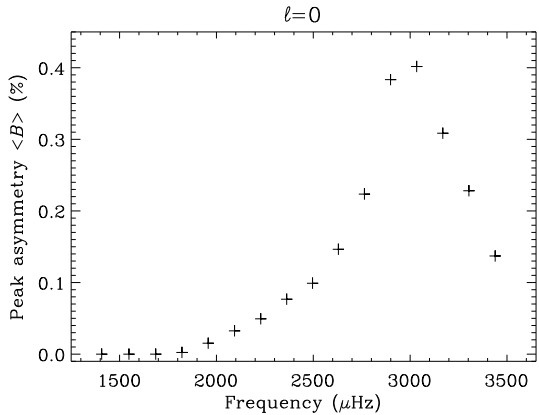}}

 \caption{Peak asymmetry, from the devil-in-the-detail effect, that
 was given to artificial $l=0$ modes in the 9.5-yr FLAG\#1
 simulations.}

 \label{fig:ass}
\end{figure}

We recall from Section~\ref{sec:prob} that the devil-in-the-detail
effect should also give rise to peak asymmetry. Fig.~\ref{fig:ass}
plots the peak asymmetry given to the $l=0$ modes in the 9.5-yr
FLAG\#1 simulations. We re-emphasize that this asymmetry comes from
the interplay of the variations in frequency, power and damping over
the simulated 9.5-yr period of observation.

Observations of the real low-$l$ modes show peak asymmetry that is
close to zero at frequencies of $\sim 3000\,\rm \mu Hz$. The asymmetry
increases in size to a few per cent at the extreme ends of the visible
p-mode spectrum. The asymmetry is also negative in the Doppler
velocity data, i.e., there is more power observed on the low-frequency
side of the peaks.  The data plotted in Fig.~\ref{fig:ass} provide a
guide to that part of the observed asymmetry of the real solar p modes
that might be expected to come from the devil-in-the-detail effect,
assuming the real observations to be of similar length, and to cover
similar levels of activity, as the artificial data here.

The devil-in-the-detail contribution is seen to be smallest at both
ends of the plotted frequency range, where the real observed
asymmetries happen to take their largest values. We would therefore
not expect there to be a significant bias in the asymmetries in these
parts of the spectrum. However, the observed asymmetries of the modes
near the centre of the spectrum are much smaller in magnitude, and
this is where the devil-in-the-detail asymmetry contribution has it
largest values. We might therefore expect a more significant bias in
this central part of the p-mode spectrum.

%%%%%%%%%%%%%%%%%%%%%%%%%%%%%%%%%%%%%%%%%%%%%%%%%%%%%%%%%%%%%%%%%%%%%%%

%%%%%%%%%%%%%%%%%%%%%%%%%%%%%%%%%%%%%%%%%%%%%%%%%%%%%%%%%%%%%%%%%%%%%%%

\section{Conclusions}
\label{sec:conc}

Over the course of the solar cycle, variations are observed not only
in the p-mode frequencies, but also in the p-mode powers. Variation of
the powers over time gives rise to a cross-talk effect with the
varying frequencies, which we call the devil-in-the-detail effect.

The effect, and its impact, may be summarized as follows. In the
absence of any power variation, the location in the power frequency
spectrum of the centroid of a mode peak would in principle give an
accurate measure of the unweighted average mode frequency over the
period of observation. However, when significant time variation of the
power is present, data samples from times when the mode is more
prominent (i.e., at high power) will carry proportionately larger
weights in determining the final peak profile than samples from times
when the mode is less prominent (i.e., at low power). As the level of
solar activity rises, the frequencies of the low-$l$ p modes under
study here increase, but corresponding mode powers
decrease. This means that, for a long timeseries, the frequency
of a mode peak will be biased so it is \emph{lower} than the
unweighted time-averaged frequency. This is our so-called devil in the
detail.

 \subsection{Impact on Sun-as-a-star data}
 \label{sec:concsas}

We have used extensive Monte Carlo simulations to estimate the
magnitude of the resulting devil-in-the-detail bias, in BiSON- and
GOLF-like Sun-as-a-star helioseismic data spanning almost one full
activity cycle. Our simulations imply the bias should rise
monotonically in frequency up to $\approx 3000\,\rm \mu Hz$, and then
level off, or even decrease slightly in magnitude, at higher
frequencies.  It is at approximately the same frequency that the
magnitude of the bias reaches a size that is comparable to the
expected frequency uncertainties. Should we be worried by this
devil-in-the-detail effect? Let us consider briefly, in this and the
next section, two issues where the effect might be a potential cause
for concern.

There is the impact of the effect on attempts to correct the
frequencies for solar activity. Since our in-depth analysis has been
made for Sun-as-a-star data on the low-$l$ p modes, we are in a
position to comment in detail on correction procedures for these data.
Procedures which attempt to `remove' the solar-cycle effects from the
fitted frequencies (e.g., the BiSON correction procedure; see Basu et
al. 2007) assume that the observed mode frequencies correspond to
unweighted time averages of the frequencies. The devil-in-the-detail
effect therefore introduces a bias in the correction, which will be
comparable in size to the frequency uncertainties for modes at
frequencies above $\approx 3000\,\rm \mu Hz$. Will this also be true
for analysis of resolved-Sun data on modes of higher $l$?

 \subsection{Brief discussion on $l$ dependence of effect}
 \label{sec:concl}

The issue of the $l$ dependence of the effect is of course important
where inversions for the internal structure are concerned, since those
inversions must make use of data on modes covering a wide range in
$l$. It is a potential complication not just for analysis of
non-contemporaneous data from two instruments (or more); but also for
analysis of data from a single instrument, since significant variation
of the bias with $l$ could affect the accuracy of inversions. The
impact of the bias on analysis of medium and high-$l$ data, from
resolved-Sun observations, will be covered in detail in a future
paper. Here, in order to get a feel for the likely size of the bias,
its variation with $l$, and the possible impact on structure
inversions, we conducted the following experiment.

First, we estimated the size of the devil-in-the-detail bias for
different $l$ in the range $0 \le l \le 150$, but only at fixed
frequency (here, a notional mode frequency of $3000\,\rm \mu Hz$). We
assumed the frequency shifts should scale inversely with the mode
inertias (and used the model `S' inertias; see Christensen-Dalsgaard
et al. 1996); and we estimated the $l$-dependent mode height and
linewidth changes from GONG data analyzed in Komm, Howe \& Hill (2000)
[their Figs.~10 and 11). With estimates of the solar-cycle frequency,
height and linewidth changes to hand for each $l$, we were in a
position to use Equation~\ref{eq:int} to construct the expected final
peak profiles -- assuming a timeseries of length 9.5\,yr (as in
Sections~\ref{sec:data} and~\ref{sec:full}) -- and to thereby estimate
the devil-in-the-detail bias as a function of $l$. Our results showed
little variation of the estimated bias with $l$ below $l \la 60$ (it
being $\approx 0.02\,\rm \mu Hz$ in size). However, at higher degrees
the bias increased in a linear manner with increasing $l$, reaching a
size of $\approx 0.05\,\rm \mu Hz$ at $l \sim 150$.

Next, we scaled the estimated bias for each $l$ by the inertia ratio
of Christensen-Dalsgaard \& Berthomieu (1991), and compared these
scaled differences with sizes of scaled frequency differences that are
typically large enough to affect inversion results for the internal
structure at the $1\sigma$ level (see also Eff~Darwich et
al. 2002). This assumed that the observed p-mode frequencies came from
the MDI frequency set of Korzennik (2005), which includes frequencies
on modes from $0 \le l \le 125$. The results of this comparison
suggest that the devil-in-the-detail bias is not quite large enough to
affect the inversions at the $1\sigma$ level, although the amounts by
which it fell short were marginal.

So, our preliminary conclusion, regarding the impact of the $l$
dependence, is that the devil in the detail may not be a cause for
concern where inversions for structure using data from a single
instrument are concerned. That said, we should stress that this
comparative analysis was an approximate one. At no point did we
account for any possible $m$ dependence, which might have a
significant influence on the results. Without having done a full Monte
Carlo simulation of the data, we were not in a position to provide an
accurate estimate of the frequency dependence for different $l$. These
factors will be tested in upcoming work.

\subsection*{ACKNOWLEDGMENTS}

SJJ-R, WJC, YE and RN acknowledge the support of the UK Science and
Technology Facilities Council (STFC). SJJ-R also thanks the European
Helio- and Asteroseismology (HELAS) Network for support. HELAS is a
major international collaboration funded by the European Commission's
Sixth Framework Program. The NSO/Kitt Peak data, used as part of the
artificial (solarFLAG) timeseries construction, were produced
cooperatively by NSF/NOAO, NASA/GSFC and NOAA/SEL. We thank the
solarFLAG collaboration for use of its artificial data.

\appendix

\section{Variation of artificial mode parameters}
\label{sec:vars}

A useful model for the solar p modes is a forced, damped harmonic
oscillator. The artificial data described in the paper are based on
oscillators described by the equation:
\begin{equation}
\ddot{x}(t) + 2\eta(t)\,\dot{x}(t) + [2\pi\nu(t)]^2 x(t) = K(t)\,
\delta(t-t_0).
\label{eq:osc}
\end{equation}
Here: $x(t)$ is the displacement of the oscillator; $\nu(t)$ is its
cyclic frequency; $\eta(t)$ is the linear damping rate; and $K(t)$ is
the amplitude of the forcing function (the `kick'), which is drawn
from a Gaussian distribution with zero mean and variance
$\sigma_K(t)$, i.e. $K(t) \sim N \left[ 0,\sigma_K(t) \right]$.

Here, we have given the basic oscillator parameters an explicit
dependence on $t$ to indicate the parameters are varied in simulated
time.  It can be shown (e.g., Chaplin et al. 2000; see also Simoniello
et al. 2004) that the implied instantaneous maximum power spectral
density at time $t$, of a timeseries of the oscillator velocity
$\dot{x}(t)$, is described by:
\begin{equation}
H(t) \propto \frac{\sigma_K^2(t)}{\eta^2(t)}.
\label{eq:ht}
\end{equation}
In the first simulation scenario in Section~\ref{sec:prob}, where
$\sigma_K(t)$ was varied in time but $\eta(t)$ remained fixed, we
therefore have:
\begin{equation}
H(t) \propto \sigma_K^2(t),
\label{eq:scen1}
\end{equation}
so that
\begin{equation}
\Delta H / H(0) = 2 \Delta \sigma_K / \sigma_K(0).
\label{eq:scen1a}
\end{equation}
In the second simulation scenario in Section~\ref{sec:prob}, where
$\eta(t)$ was varied in time but $\sigma_K(t)$ remained fixed, we
have:
\[
H(t) \propto 1/\eta^2(t),
\]
\begin{equation}
\Gamma(t) \propto \eta(t).
\label{eq:scen2}
\end{equation}
Therefore:
\[
\Delta H/H(0) = -2 \Delta \eta / \eta(0),
\]
\begin{equation}
\Delta \Gamma / \Gamma(0) = \Delta \eta / \eta(0).
\label{eq:scen2a}
\end{equation}
So, in the second scenario both the height \emph{and} linewidth
change in time. The fractional variation in the height is twice that
in the width, and the fractional changes have opposite sign.

\end{document}